\documentclass[a4paper,%
   a4paper,%
   BCOR12mm,%
   11pt,%
   abstracton,%
   pointednumbers,%
   tablecaptionabove,%
   footinclude,%
   halfparskip,%
   ]{scrartcl}
\usepackage[ilines]{scrpage2} 
\usepackage{epsfig}
\usepackage{multicol}
\usepackage{amsmath}
\usepackage[square,comma,sort,numbers]{natbib}

\newcommand{\UPLB}{Unibersidad ng Pilipinas Los Ba\~{n}os}
\newcommand{\GRAPH}{\mathcal{G}}
\newcommand{\VERTEXSET}{\mathcal{V}}
\newcommand{\EDGESET}{\mathcal{E}}
\newcommand{\ACCOUNT}{\mathcal{A}}

\title{Ang Social Network sa Facebook ng mga\\Taga-Batangas at ng mga Taga-Laguna:\\Isang Paghahambing}
\author{Jaderick P. Pabico\footnote{Surian ng Agham Pang-kompyuter} at Jose Rene L. Micor\footnote{Surian ng Kapnayan}\\
   \UPLB\\
   College 4031, Laguna}
\date{}

\lehead{}
\rohead{\thepage}
\lofoot{}
\rofoot{}

\begin{document}
\maketitle

\begin{abstract}
{\bf [English Abstract]} Online social networking (OSN) has become of great influence to Filipinos, where Facebook, Twitter, LinkedIn, Google+, and Instagram are among the popular ones. Their popularity, coupled with their intuitive and interactive use, allow one's personal information such as gender, age, address, relationship status, and list of friends to become publicly available. The accessibility of information from these sites allow, with the aid of computers, for the study of a wide population's characteristics even in a provincial scale. 
Aside from being neighbouring locales, the respective residents of Laguna and Batangas both derive their livelihoods from two lakes, Laguna de Bay and Taal Lake. Both residents experience similar problems, such as that, among many others, of fish kill. The goal of this research is to find out similarities in their respective online populations, particularly that of Facebook's. With the use of computational dynamic social network analysis (CDSNA), we found out that the two communities are similar, among others, as follows:
\begin{enumerate}
\item Both populations are dominated by single young female;
\item Homophily was observed when choosing a friend in terms of age (i.e., friendships were created more often between people whose ages do not differ by at most five years); and
\item Heterophily was observed when choosing friends in terms of gender (i.e., more friendships were created between a male and a female than between both people of the same gender).
\end{enumerate}
This paper also presents the differences in the structure of the two social networks, such as degrees of separation and preferential attachment.

{\bf [Filipino Abstract]} Ang online social networking (OSN) sa Internet ay kasalukuyang may malaganap na impluwensiya sa buhay ng mga Pilipino, kung saan ang Facebook, Twitter, LinkedIn, Google+, at Instagram ay ilan sa mga sikat. Ang kasikatan at pagiging madaling magamit ng mga OSN ay nagbigay daan para ang personal na impormasyon katulad ng kasarian, gulang, pook na tinitirahan, estadong pang-sibil at mga listahan ng mga kaibigan ay maging pampublikong kaalaman. Dahil dito, sa tulong ng paggamit ng kompyuter, napapadali ang pag-aaral sa mga katangian ng mga populasyon sa mas malawak na iskala, kahit kasing lawak ng mga lalawigan.

Bukod sa sila ay magkalapit pook, ang mga mamayan ng dalawang lalawigan ng Laguna at ng Batangas ay parehong nabibigyang buhay ng dalawang lawa, ang mga lawa ng Laguna de Bay at ng Taal. Magkatulad din ang kanilang mga suliranin, halimba sa {\em fish kill}. Nais malaman ng pagsasaliksik na ito kung ang kanilang mga populasyon sa Internet, partikular na ang Facebook, ay magkahambing din. Sa pamamagitan ng {\em computational dynamic social network analysis} (CDSNA), napag-alaman na ang dalawang pamayanan ay magkatulad sa mga sumusunod (bukod sa iba pa):
\begin{enumerate}
	\item	Ang populasyon ay dominado ng mga kabataang babae na walang asawa (o ka-relasyon);
	\item	Kapansin-pansin ang {\em homophily} sa pagpili ng ka-{\em friend} sa pamamagitan ng gulang (mas madali silang maging magkaibigan kung hindi mas mataas mula sa lima hanggang sampung taon ang agwat ng mga gulang nila); at
	\item	{\em Heterophily} naman ang mapapansin sa pagpili ng ka-{\em friend} kung ang pagbabasehan ay ang kasarian (mas maraming relasyon ang nabubuo sa pagitan ng isang babae at isang lalaki kesa sa pagitan ng dalawang babae o dalawang lalaki).
\end{enumerate}
Ilalahad din ng diskursong ito ang pagkakaiba sa istraktura ng dalawang {\em social network}, kabilang na ang mga sukat nito katulad halimbawa ng {\em degree of separation} at {\em preferential attachment}.
\end{abstract}


\section{Panimula}\label{intro}
Sa simula ng ika-21 siglo, ang malaganap na katangian ng {\em Internet} ay naging mahalaga sa buhay ng mga konektadong Pilipino sa pamamagitan ng iba't-ibang mga serbisyo sa {\em web} na tinaguriang {\em online social networking} (OSN). Ang mga halimbawa ng OSN na naging sikat sa mga Pilipino ay ang Facebook\footnote{\tt http://www.facebook.com}, Twitter\footnote{\tt http://www.tiwtter.com}, LinkedIn\footnote{\tt http://www.linkedin.com}, Google+\footnote{\tt http://plus.google.com}, at Instagram\footnote{\tt http://www.instagram.com}, na kung saan 6.9~milyon sa 7.9~milyong mga Pilipino ang bumibisita kada buwan~\citep{yazon07,liao08,wikipedia08}. Sa katunayan, isang pag-aaral~\citep{mccann08} ang nagdeklarang ang Pilipinas ang kapitolyo ng social networking sa mundo, kung saan 83\% sa  mga Pilipinong sinuri ay mga kasapi ng OSN. Ang mga OSN na ito ay nagpapahintulot sa mga gumagamit na ilathala ang kanilang mga pang-personal na impormasyon katulad ng gulang, kasarian, estadong pang-sibil, pook tirahan, at maging ang listahan ng mga kaibigan. Dahil dito, ang mga OSN ay nagbigay ng daan upang makapagbuo ng bagong metodolohiya sa pagsasaliksik at mapag-aralan ang demograpya sa malawakang iskala na hindi kailangang gumamit ng tradisyonal na metodolohiya katulad ng {\em surveying}.


Sa mga nagdaang mga taon, marami ng mga pagsusuri ang nakapagtala at nakapagbigay linaw sa mga istraktura ng mga {\em virtual community} (VC). Tatlong pag-aaral na maari nating gawing halimbawa ay ang kay Zinoviev~\citep{zinoviev08},  kina Leskovec at Horvitz~\citep{leskovec08}, at kina Pabico at Arevalo~\citep{pabico08}. Si Zinoviev ay sumubok na alamin ang topolohiya at heometrya ng {\em Moi Krug}\footnote{http://moikrug.ru}, isang OSN sa bansang Rusya. Gumamit siya ng iba't-ibang matematikal na pangsukat sa isang {\em graph}, katulad ng {\em vertex degree} at {\em path length}. Sa kanyang pagsisikap na bigyang liwanag ang istraktura, hangganan, at panloob na lugar ng komunidad na ito, kanyang nilikha ang mga konsepto ng {\em dense core} at {\em local density} na kanyang ginamit para makilala ang mga sikat (o may malakas na impluwensiya) at di-sikat (o walang impluwensiya) sa komunidad. Samantala, sina Leskovec at Horvitz naman ay nagsuri ng mga katangian at mga nabuong anyo na galing sa sama-samang pagkilos ng mga kasapi ng isang malaking VC. Ang mga kasapi dito ay lumahok sa isang mataas na antas ng komunikasyon sa pamamagitan ng {\em Microsoft instant-messaging system}. Galing sa kanilang datos, kanilang binuo ang tinatawag na {\em communication network} (CN) na may 180 milyong kasapi at may 1.3 bilyong mga pag-uusap. Kanilang napag-alaman na ang {\em average degree of separation} ng CN ay sumasang-ayon sa {\em six degrees of separation} nina Travers at Milgram~\citep{travers69}. 

Sa mga pagsasaliksik nina Pabico at Arevalo~\citep{pabico08} sa noo'y tanyag na Friendster OSN, gumawa sila ng {\em web spider}, isang kompyuter {\em program}, na awtomatikong nangngalap ng datos ng 7,172 na mga kasapi galing sa Los Ba\~nos, Laguna. Kanilang napag-alaman na: (1)~Mas maraming mga babaeng kasapi (52.34\%) keysa sa mga lalake (47.66\%); (2)~Ang mga kasapi na may mga gulang na 15--25 na taon ng parehong kasarian ang bumuo ng 68\% ng OSN, kasunod ng mga may gulang na 26--40 sa 28\%, at mga may gulang ng 41--85 sa 4\%; Ang mga {\em senior citizen} na may gulang na 64--85 ay bumuo sa 1\%; (3)~{\em Homophily}, o ang kasabihang {\em birds-of-a-feather}, ay napansin sa pagpili  ng kaibigan kung ang pagbabatayan ay ang gulang, o ang mga kasapi ay may pagkiling na maging kaibigan ang kasing-gulang nila; (4)~Napansin naman ang {\em heterophily} sa pagpili ng mga kaibigan kung ang pagbabatayan ay kasarian lamang; (5)~Ang Friendster OSN ay nagpakita ng katangiang {\em small world} dahil sa ang {\em average path length} na 4.5 nito ay mas maliit keysa sa tanyag na ``{\em six degrees of separation}'' na natuklasan nina Travers at Milgram~\citep{travers69}; At, (6)~ang OSN ay nagtatanghal ng katangiang {\em scale free} na nagpapahiwatig ng pagkakaroon ng mga kasaping gumaganap bilang sentro ng komunidad.

Sa pagsasaliksik na ito, ginamit namin ang {\em web spider} nina Pabico at Arevalo~\citep{pabico08} para kolektahin ang mga datos ng mga kasapi sa Facebook na galing sa dalawang lalawigang magkalapit, ang mga lalawigan ng Laguna at ng Batangas. Sa kasalukuyan, ang Facebook ang pinakatanyag na OSN sa buong mundo na may 1.11 bilyon\footnote{http://news.yahoo.com/number-active-users-facebook-over-230449748.html} na mga kasapi, 22.5 milyon\footnote{http://www.socialbakers.com/facebook-statistics/philippines} nito ay mga Pilipino. Ang mga mamayan ng dalawang lalawigang ito ay parehong pinili dahil karamihan sa kanilang kabuhayan ay umaasa sa mga lawa ng Laguna de Bay at ng Taal. Dahil dito, masasabing magkatulad ang kanilang mga kultura na parehong batay sa mga kultura ng pangisdaan. Isang halimbawa na lang ay ang magkatulad nilang suliranin sa {\em fish kill} na kapag dumarating sa kanilang mga buhay, sila ay may magkahalintulad na mekanismo para malampasan ito~\citep{macandog12}. Dahil dito, nais din naming malaman na kung pagkakatulad ang kanilang mga pamayanan, sila din kaya ay magkatulad sa larangan ng {\em social network}?

Ang {\em web spider} (o {\em web robot} sa ibang teksto) ay aming ipinasadya para gamitin lamang sa Facebook. Eto ay gumapang sa {\em web} ng Facebook para magtipon ng mga datos tungkol sa mga kasapi ng Facebook na nagmula sa Batangas o sa Laguna. Ang mga datos na tinipon nito ay ang kasarian, gulang, estadong pang-sibil, at mga listahan ng kaibigan. Ginamit namin ang {\em undirected graph}~$\GRAPH(\VERTEXSET, \EDGESET)$ para maisalarawan sa matematikong paraan ang mga relasyon ng bawat kasapi, na kung saan ang~$\VERTEXSET$ ay sumasagisag sa kalipunan ng mga {\em vertex} $\{v_1, v_2, \dots, v_n\}$ at ang~$\EDGESET$ naman ay kumakatawan sa pangkat ng mga {\em edge} $\{(i,j)|v_i, v_j\in\VERTEXSET\}$. Ang lahat ng {\em vertex}~$v_1, v_2, \dots, v_n$ ay sumasalamin sa mga taong kasapi ng OSN, samantalang ang mga {\em edge} naman ay tumuturing sa mga relasyon nila. Halimbawa, ang {\em edge} na~$(i,j)$ ay sumisimbolo na may kaugnayan o relasyon sina~$v_i$ at~$v_j$. Ang ibig sabihin nito, ang {\em edge} na $(i,j)$ ay nagsasaad na sina~$v_i$ at~$v_j$ ay magkaibigan (o ka-{\em friend} sa linguwahe ng OSN).

\section{Ang {\em Facebook Database}}\label{facebook}

\subsection{Pagtipon ng Datos sa {\em Facebook}}\label{extract}
Ang Facebook ay isang OSN na itinatag nina Mark Zuckerberg, Eduardo Saverin, Andrew McCollum, Dustin Moskovitz at Chris Hughes~\citep{carlson10}. Ang Facebook ay halaw sa mga pamamaraang {\em Circle of Friends} at {\em Web of Friends} na nag-uugnay sa mga tao sa {\em virtual} na pamayanan~\cite{rosen07}. Sa kasalukuyan, ang Facebook ay may 1.11 bilyong {\em account} na tumutumbas sa bilang ng mga taong kasapi nito sa buong mundo. 

Ang Facebook ay gumagamit ng mga metodolohiya at awtomatikong pamamaraan sa pamamagitan ng kompyuter para matuos, maipakita, at magsagawa ng aksyon sa mga relasyon ng mga kasapi nito. Pinaghalo nito ang mga awtomatikong pamamaraan na tinaguriang {\em Web of Friends} at ang mga pamamaraan sa tinaguriang {\em Web of Contacts} para magtipon ng mga datos na nagbibigay anyo sa mga kasapi nito. Ang pamamaraan ay nagpapahintulot sa mga kasapi ng Facebook para malagyan ng tanda (o {\em tag}) ang iba pang mga kasapi nito na kanilang mga kaibigan, ka-relasyon, o ka-{\em friend}. Halimbawa, kung sina~$v_i$ at~$v_j$ ay magkaibigan sa tutuong buhay, at pareho silang kasapi sa Facebook, maaring i-{\em tag} ni~$v_i$ si~$v_j$ para malaman ng Facebook na sila ay magkaibigan. Kung magkaganoon, lahat ng mga kaibigan ni~$v_i$ ay malalaman ni~$v_j$, na maari ring i-{\em tag} ni~$v_j$ para maging kabigan din nila sa OSN. Sa pamamagitan ng metodolohiyang ito, lahat ng personal na datos, kabilang na ang mga datos ng mga kaibigan, ay naisasama-sama at sistematikong napro-proseso para magbigay linaw sa mga tagni-tagning relasyon na nagkokonekta sa kahit sinong dalawang kasapi ng Facebook sa buong mundo. Kaya, mula sa isang kasapi, maaring malaman ng kahit sino ang pinakamalapit na kakilala para marating o maka-konekta sa isa pang kasapi na malayo sa kanya o hindi siya kilala. Itong pamamaraang ito ay nagsasalamin sa {\em padrino system} na nakatanim na sa kulturang Pilipino. Halimbawa, kung si~$v_i$ ay may mahalagang kailagan kay~$v_k$ at kapwa sila walang tahas na koneksyon na namamagitan sa kanila, at kung si~$v_i$ ay kaibigan ni~$v_j$, at si $v_j$ naman ay kaibigan ni~$v_k$, kung magkagayo'y maaring mag-{\em tulay} kina~$v_i$ at~$v_k$ si~$v_j$. Dito, masasabi nating naging {\em padrino} ni~$v_i$ si~$v_j$ para makilala ni~$v_k$. Ang mga metodologhiyang ito ay mas pinadali pa ng mga makabagong gamit pangkomunikasyong--OSN na ibinibigay bilang libreng serbisyo ng Facebook.
\subsection{Ang {\em Web Robot}}

Ang {\em web robot} ay ginawa sa pamamagitan ng {\em perl script} at pinatakbo sa kompyuter sa pamamagitan ng mga {\em Linux command line utility program} na {\tt grep}~\citep{grep} at {\tt wget}~\citep{wget}. Isang Facebook {\em account}~$\ACCOUNT$ ang amin munang unang ginawa dahil mga naka-{\em logon} na mga kasapi lang ang maaring makakita ng mga pang-personal na datos ng iba pang mga kasapi. Dito, ang~$\ACCOUNT$ ay sa katunayang {\em account} ng isang kasapi sa Facebook at hindi isang {\em dummy account} lamang. Ang Facebook ay may palagian ng takdang paglinis mula sa kanilang OSN ng mga {\em Pretendsters}, {\em Fakesters}, at {\em Fraudsters}~\citep{terdiman04}, kaya mayroong pagpapatunay na ang mga datos na makakalap ng {\em web robot} ay mga datos ng tutuo at buhay na mga tao. Ang {\em web robot} ay gumamit ng {\em cookie file} sa {\em web browser} ng kasalukuyang naka-{\em logged-in} na si~$\ACCOUNT$. Kaya, sa palagay ng Facebook, ang {\em web robot} ay walang anuman kundi si $\ACCOUNT$~mismo.

Ang {\em web robot} ay nagkalap ng mga datos sa Facebook mula ika-11 ng Hulyo hanggang ika-6 ng Agosto ng 2013. Gamit ang {\em cluster} sa Surian ng Agham Pang-kompyuter sa \UPLB\ na binubuo ng 16~na mga kompyuter na may {\em operating system} na {\em Scientific Linux version 6.4}, ang {\em web robot} ay nagtala ng mga datos sa isang {\em MySQL database}. Dahil may 16~kompyuter ang {\em cluster}, ibig sabihin ay mayroon ding 16~na sabay-sabay at nagtutulungang mga {\em web robot} ang gumawa ng trabaho. Mas matagal kasing makalap ang mga datos kung mas maliit na bilang ng kompyuter ang gagamitin. Dahil sa mas mabilis magdagdag at magtanggal ng kasapi ang Facebook, kailangang mas mabilis ding makalap ng mga {\em web robot} ang mga datos. Ayon sa teorya~\citep{pabico14}, mas mabilis makakalap ang mga datos bago pa man lang magbago ang OSN kung mas madaming bilang ng kompyuter ang sabay-sabay na gagamitin. Subalit dahil sa pagsasaalang-alang sa mga pang-pinansiyal na bagay at sa hangganan ng mga pisikal na sistema, hanggang 16~lamang ang aming nagamit. Dahil dito, ipinagpalagay namin na ang mga pagbabagong naganap sa OSN sa loob ng mga araw ng aming pagkalap ng datos ay walang pagkakaibang estadistika\footnote{Maaring mapabulaanan ang pagpapalagay na ito kung sa darating na mga panahon ay malalampasan ng mga pisikal na sistema ang kasalukuyang mga hangganan nito.}.

\subsection{Ang OSN ng mga taga-Batangas at ng mga taga-Laguna}

Ang {\em search tool} ng Facebook ay aming ginamit para maitala ang mga {\em account} ng mga kasaping naglathala ng Batangas o ng Laguna bilang kanilang mga pook--tirahan. Ang {\em search parameter} na aming ginamit ay yaong magtatala ng lahat ng kasarian, may pinakamalawak na saklaw ng gulang, at mga datos na nagsasaad sa estadong pang-sibil. Bukod sa may asawa at walang asawa, ang Facebook ay may karadagang estadong pang-sibil na amin ding sinundan, ang ``{\em in a relationship}'' o IAR, at ang ``{\em unknown}.'' Ang IAR ay nagsaaad sa isang kasapi, na maaring may asawa o walang asawa, sa kanyang kasalukuyang estadong pang-relasyon. Halimbawa, maaring wala siyang asawa subalit meron siyang kasintahan. O dili naman kaya, maaari siyang may asawa subalit hiwalay o patay na at kasalukuyan nakikipagrelasyon siya sa ibang tao. Ang {\em unknown} naman ay mga kasaping ayaw maglahad ng kanilang mga estadong pangrelasyon.

Ang {\em search tool} ay nagpakita ng isang {\em array} $L=\{l_0, l_1, \dots, l_{p-1}\}$ ng $p$~mga {\em web pages} na bawat isa nito ay nagtala ng~$N$ na mga {\em accounts}. Ang bawat isa sa mga {\em web page} na $l_0, l_1, \dots, l_{p-2}$ ay nagtala ng sampung {\em account}, samantalang ang huling {\em web page}~$l_{p-1}$ ay nagtala ng $p$~{\em modulo}~$N$ na mga {\em account}. Ang {\em web robot} ay nagsimulang mag-{\em crawl} sa~$l_0$ at ginamit nito ang URL\footnote{Uniform Resource Locator} para ma-{\em crawl} din ang mga~$l_i$, $\forall i=1,\dots,p-1$, na nagsasaad ng listahan ng mga kaibigan ni~$l_0$. Sa bawat {\em web page}~$l_i$, ang {\em web robot} ay awtomatikong nagtala ng {\em account number}, {\em user name}, gulang, kasarian, at estadong pang-sibil ng bawat kasapi. Ang mga natipong datos ay inilagay sa isang {\em database}~$\Delta_u$. Sa katapusan ng pag-{\em crawl} ng {\em web robot}, ang~$\Delta_u$ ay nagkaroon ng~$N$ na mga tala na nag-uugnay sa $N$~na mga kasapi ng Facebook na taga-Batangas o taga-Laguna.

Habang ang {\em web robot} ay nagka-{\em crawl}, ang mga listahan ng mga kaibigan ng bawat kasapi ay dinaanan din ng {\em web robot} at ang kanilang mga datos ay kinuha at itinala sa~$\Delta_u$. Subalit ang listahan ng mga kaibigan ay itinala sa hiwalay na {\em database}~$\Delta_f$. Ang mga {\em database} na~$\Delta_u$ at~$\Delta_f$ ay mayroong {\em one--to--many} na relasyon.
\subsection{Pagtuos sa Datos ng Demograpiya}

Ang mga sumusunod na datos ng demograpiya ay natuos galing sa~$\Delta_u$:
\begin{enumerate}
\item Bilang~$N_g$ at bahagdan~$P_g$ ng mga kasapi, base sa kasarian~$\delta_g$;
\item Bilang~$N_a$ at bahagdan~$P_a$ ng mga kasapi, base sa gulang~$\delta_a$;
\item Bilang~$N_r$ at bahagdan~$P_r$ ng mga kasapi, base sa estadong pang-sibil~$\delta_r$;
\item Bilang~$N_{g\times a}$ at bahagdan~$P_{g\times a}$ ng mga kasapi, base sa~$\delta_g$ at~$\delta_a$;
\item Bilang~$N_{g\times r}$ at bahagdan~$P_{g\times r}$ ng mga kasapi, base sa~$\delta_g$ at~$\delta_r$;
\item Bilang~$N_{a\times r}$ at bahagdan~$P_{a\times r}$ ng mga kasapi, base sa~$\delta_a$ at~$\delta_r$; at
\item Bilang~$N_{g\times a\times r}$ at bahagdan~$P_{g\times a\times r}$ ng mga kasapi, base sa~$\delta_g$, $\delta_a$ at~$\delta_r$.
\end{enumerate}
Dito, madaling makita na ang mga estadistikang datos, katulad ng mga bahagdan ay matutuos sa pamamagitan ng mga bilang, ayon sa mga sumusunod:
\begin{eqnarray}
  P_g &=& \frac{N_g}{N}\times 100\%;\nonumber\\
  P_a &=& \frac{N_a}{N}\times 100\%;\nonumber\\
  P_r &=& \frac{N_r}{N}\times 100\%;\nonumber\\
  P_{g\times a} &=& \frac{N_{g\times a}}{N}\times 100\%;\nonumber\\
  P_{g\times r} &=& \frac{N_{g\times r}}{N}\times 100\%;\nonumber\\
  P_{a\times r} &=& \frac{N_{a\times r}}{N}\times 100\%; {\rm at}\nonumber\\
  P_{g\times a\times r} &=& \frac{N_{g\times a\times r}}{N}\times 100\%.\nonumber
\end{eqnarray}
Ang mga batayang estadistika ng bilang ay isinasalarawan ng mga sumusunod:
\begin{enumerate}
  \item $N_g$ ay alinman sa $N_{\rm male}$ o $N_{\rm female}$. Tandaan na $N_{\rm male} + N_{\rm female} = N$.
  \item $N_a$ ay isa sa mga $N_{\rm 18}$, $N_{\rm 19}$, $\dots$, o $N_{\rm 65}$. Tandaan dito na $\sum_{i=18}^{65}N_i=N$. Dito, ang mga {\em subscript} ay nagsasaad ng taong gulang.
  \item $N_r$ ay alinman sa $N_{\rm single}$, $N_{\rm married}$, $N_{\rm IAR}$, or $N_{\rm unk}$, kung saan ang IAR ay nagsasaad ng {\em in a relationship} at ang unk o {\em unknown} ay nagsasaad ng relasyong ayaw ipaalam ng may-ari. Muli, tandaan na $N_{\rm single}+N_{\rm married}+N_{\rm IAR}+N_{\rm unk}=N$.
\end{enumerate}

Ang mga pinagsamang estadistika ng bilang ay walang iba kundi ang mga bilang ng mga kasapi sa mga pinagtambal na mga datos. Halimbawa, ang bilang ng walang asawang mga lalaki na 25 taong gulang ay matutuos ng $N_{\rm male \times 25 \times single} = |\delta_{\rm male} \bigcap \delta_{\rm 25} \bigcap \delta_{\rm single}|$.
\subsection{Pag-alam sa Pagpili ng ka-{\em Friend}}

Ang mga sumusunod na pagtatangi ng mga kasapi ay napag-alaman sa pamamagitan ng~$\Delta_u$ at~$\Delta_f$:
\begin{enumerate}
\item Pagtatangi sa pamamagitan ng kasarian;
\item Pagtatangi sa pamamagitan ng gulang; at
\item Pagtatangi sa pamamagitan ng estadong pang-sibil.
\end{enumerate}

Para masuri kung mayroong anyo ng pagpili base sa kasarian, ang mga bilang ng mga pagkakaibigan sa pagitan ng dalawang lalaki, dalawang babae, at ng isang lalaki at ng isang babae ay tinuos mula sa pinagsamang mga tala sa~$\Delta_u$ at~$\Delta_f$. Gayon din, ang anyo ng mga pagtatangi sa pamamagitan ng pangkat ng gulang, gayon din sa estadong pang-sibil, ay tinuos mula sa pinagsamang talaan ng~$\Delta_u$ at~$\Delta_f$.

\subsection{Pagbuo ng {\em Graph} ng OSN}
Ang {\em graph} ng OSN ay binuo sa pamamagitan ng mga datos sa~$\Delta_f$. Gamit ang {\em account} bilang {\em vertex}, at ang mga relasyon sa pagitan ng dalawang {\em account} bilang {\em edge}, isang $N \times N$ na {\em adjacency matrix}~$R$ ang binuo. Ang mga elementong $r_{i,j}\in R$ ay may katapat na halaga na~1 kung may relasyon sa pagitan nina~$i$ at~$j$ na nakatala sa~$\Delta_f$. Kung wala naman, $r_{i,j}=0$.

Mula sa~$R$, mapapag-alaman ang mga sumusunod na pagsusukat sa OSN:
\begin{enumerate}
\item Ang pinakamaliit, pamantayan, at pinakamalaking bilang ng mga ka-relasyon ng bawat kasapi;
\item Ang pinakamaiksi, pamantayan, at pinakamahabang {\em path length} sa pagitan ng kung sino mang dalawang kasapi; at
\item Ang {\em degree distribution} ng bilang ng mga ka-relasyon na meron ang isang kasapi.
\end{enumerate}

\section{Resulta at Talakayan}\label{results}
Ayon sa 2010 {\em Census of Population and Housing} na isinagawa ng {\em National Census and Statistics Office}~\citep{NCSO10}, ang Batangas ay may populasyong 2,377,395 samantalang ang Laguna ay may 2,669,847 ({\em Table}~\ref{tab:fact}). Mula sa mga datos na nakalap ng {\em web robot}, 316,123 na mga kasapi ang nagsaad na Laguna ang kanilang pook--tirahan samantalang 234,052 naman ang sa Batangas. Kung isasaalang-alang ang nalimbag na {\em national population growth rate} na 2.18\% kada taon~\citep{NCSO10}, ang mga kasapi ng Facebook mula sa Laguna ay bumubuo ng 11\% ng tinatayang populasyon ngayong 2013, samantalang ang sa Batangas ay 9\% naman.

\begin{table}[htb]
\caption{Ang paghahambing ng mga populasyon ng Batangas at ng Laguna ayon sa bilang ng NCSO noong 2010, sa pagtataya ngayong 2013 sa pamamagitan ng 2.13\% na paglago bawat taon, at sa Facebook.}\label{tab:fact}
\centering\begin{tabular}{lcccc}
\hline\hline
Lalawigan & Populasyon & Pagtataya & Facebook & Bahagdan ng \\
          & (2010)     & (2013)    &          & Facebook\\
\hline
Batangas & 2,377,395 & 2,536,291 & 234,052 & 9\%\\
Laguna   & 2,669,847 & 2,787,522 & 316,123 & 10\%\\
\hline\hline
\end{tabular}
\end{table}
\subsection{Demograpiya}

\begin{figure}[hbt]
\centering\epsfig{file=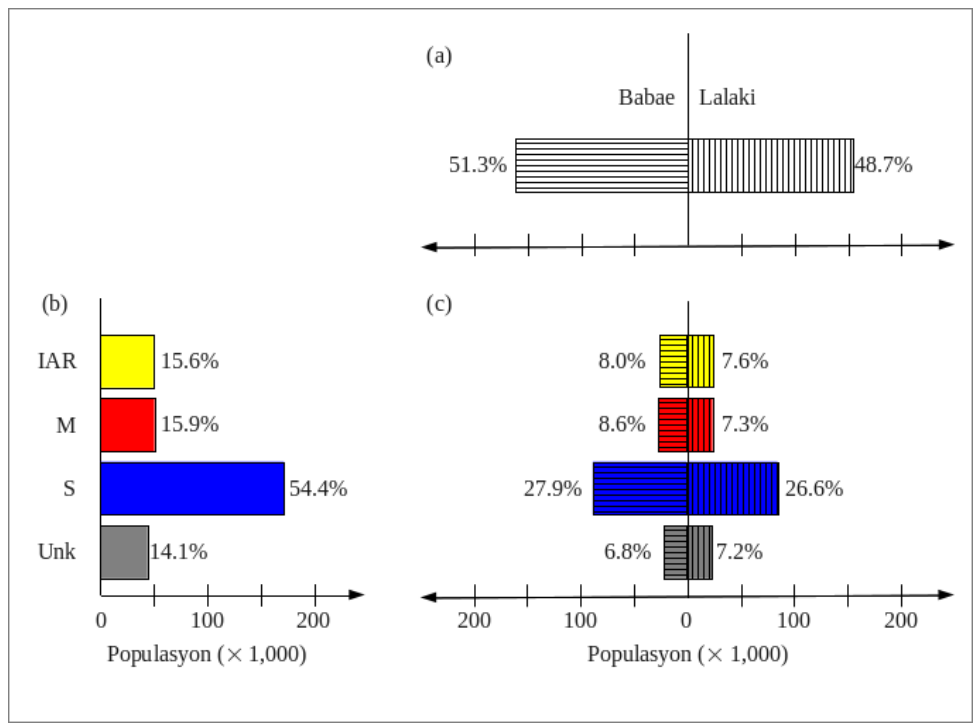, width=5in}
\caption{Bilang at bahagdan ng populasyon ng Facebook mula sa Laguna sa pamamagitan ng (a)~kasarian, (b)~estadong pang-sibil, at (c)~kasarian at estadong pang-sibil.}\label{fig:lag-demography}
\end{figure}

Ipinapakita ng {\em Figure}~\ref{fig:lag-demography} ang bilang at bahagdan ng mga kasapi ng Facebook mula sa Laguna sa pamamagitan ng (a)~kasarian, (b)~estadong pang-sibil, at (c)~ang pinagsamang kasarian at estadong pang-sibil. Sa {\em Figure}~\ref{fig:lag-demography}a, ang mga babaeng taga-Laguna ay mas nakakarami (51.3\%) kaysa sa mga lalaki (48.7\%). Ayon sa {\em Figure}~\ref{fig:lag-demography}b, ang mga walang asawa ang nangibabaw na may 54.4\%, samantalang ang naglahad ng may asawa, IAR at {\em unknown} ay, ayun sa pagkakasunud-sunod, may 15.9\%, 15.6\%, at 14.1\%. Sa {\em Figure}~\ref{fig:lag-demography}c, ang mga babaeng may asawa (27.9\%), walang asawa (8.6\%), at IAR (8\%) ay mas marami kesa sa mga lalaki (na may 26.6\%, 7.3\%, at 7.6\%, ayun sa pagkakasunud-sunod), subalit mas marami ang {\em unknown} na mga lalaki (7.2\%) kesa sa mga babae (6.8\%). Dito, makikita nating sa mga matapat na naglahad ng kanilang estadong pang-sibil, mas marami ang mga babaeng taga-Laguna kesa sa mga lalaki, samantalang sa mga hindi naglahad, mas marami ang mga lalaki.

\begin{figure}[hbt]
\centering\epsfig{file=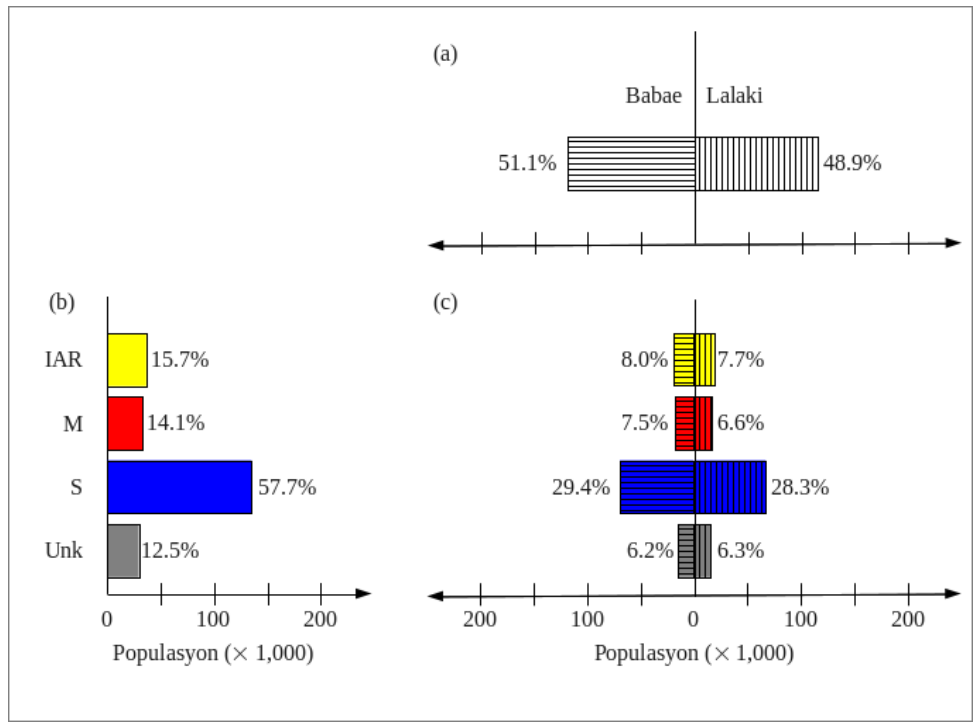, width=5in}
\caption{Bilang at bahagdan ng populasyon ng Facebook mula sa Batangas sa pamamagitan ng (a)~kasarian, (b)~estadong pang-sibil, at (c)~kasarian at estadong pang-sibil.}\label{fig:bat-demography}
\end{figure}

Ipinapakita naman ng {\em Figure}~\ref{fig:bat-demography} ang bilang at bahagdan ng mga kasapi ng Facebook mula sa Batangas. Ayon sa {\em Figure}~\ref{fig:bat-demography}a, katulad ng sa Laguna, mas marami ang mga babaeng taga-Batangas na bumubuo ng 51.1\% kesa sa mga lalaki na may 48.9\%. Ang {\em Figure}~\ref{fig:bat-demography}b naman ay nagpapakita na mas nangingibabaw ang mga walang asawa (57.7\%) kesa sa IAR (15.7\%), may asawa (14.1\%), at {\em unknown} (12.5\%). Makikita naman sa {\em Figure}~\ref{fig:bat-demography}c ang mga datos na kasing-tulad ng sa Laguna: Mas marami ang mga babaeng walang asawa (29.4\%), IAR (8\%), at may asawa (7.5\%) kesa sa mga lalaking katapat nila (28.3\%, 7.7\%, at 6.6\%, ayun sa pagkakasunud-sunod), at mas marami naman ang mga lalaking {\em unknown} (6.3\%) kesa sa mga babae (6.2\%). Muli, katulad ng mga obserbasyon natin sa taga-Laguna, marami ang mga babaeng taga-Batangas ang naglahad ng kanilang estadong pang-relasyon, samantalang mas marami naman ang mga lalaking hindi naglahad.

\begin{figure}[hbt]
\centering\epsfig{file=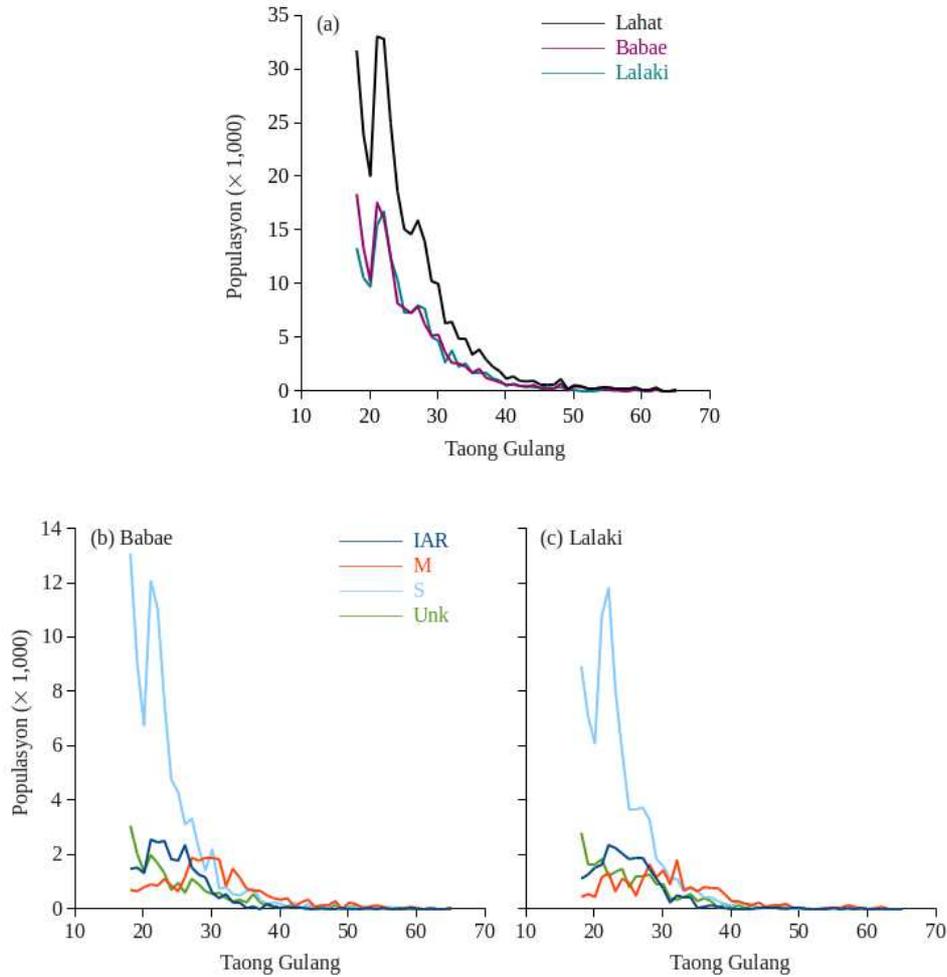, width=5in}
\caption{Bilang at bahagdan ng populasyon ng Facebook mula sa Laguna sa pamamagitan ng (a)~kasarian at gulang, (b)~estadong pang-sibil at gulang (para sa mga babae), at (c)~estadong pang-sibil at gulang (para sa mga lalaki).}\label{fig:lag-age}
\end{figure}

\begin{figure}[hbt]
\centering\epsfig{file=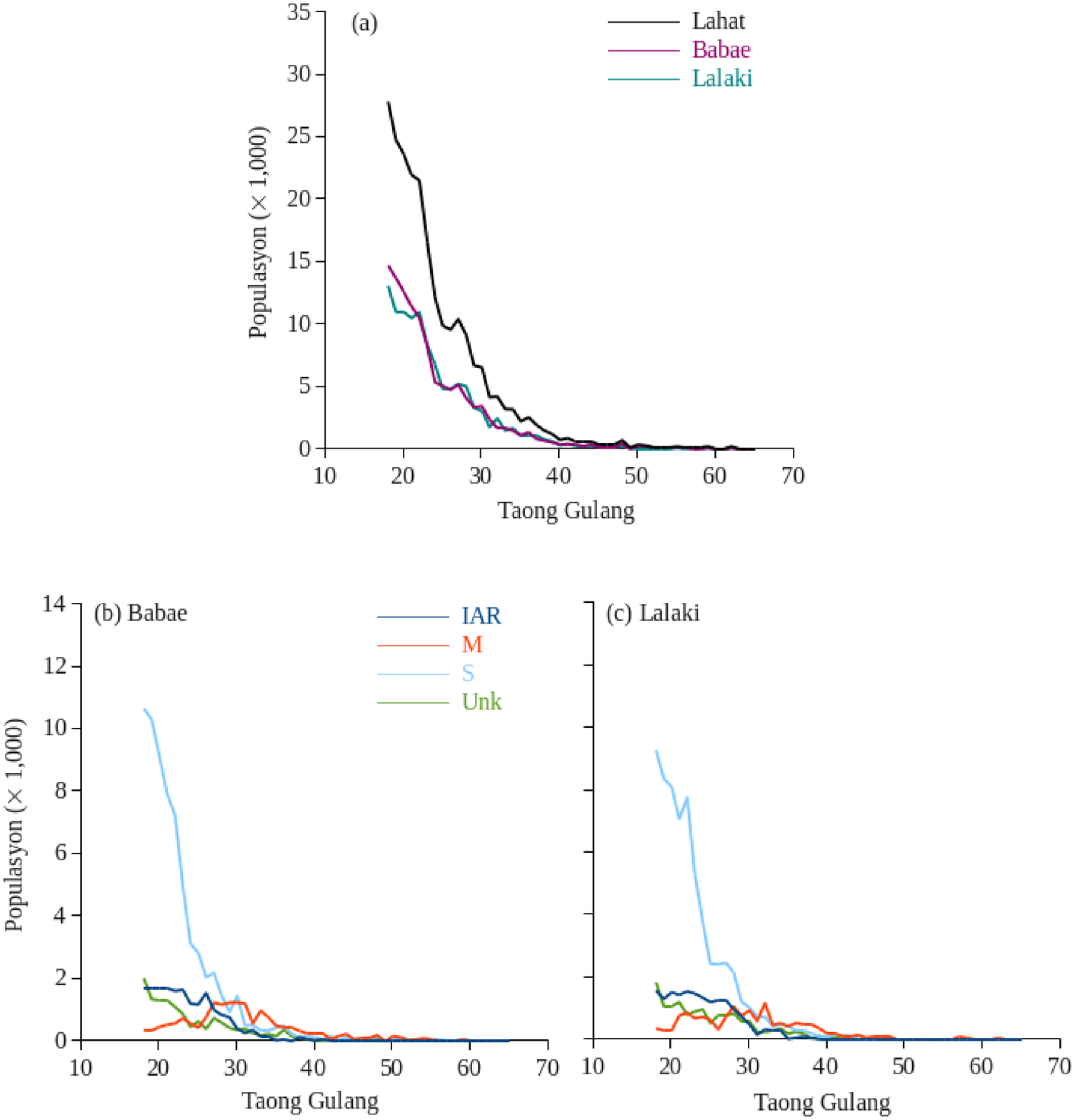, width=5in}
\caption{Bilang at bahagdan ng populasyon ng Facebook mula sa Batangas sa pamamagitan ng (a)~kasarian at gulang, (b)~estadong pang-sibil at gulang (para sa mga babae), at (c)~estadong pang-sibil at gulang (para sa mga lalaki).}\label{fig:bat-age}
\end{figure}

Ang mga sukat demograpiya ayun sa gulang ay ipinapakita sa {\em Figure}~\ref{fig:lag-age} para sa taga-Laguna at sa {\em Figure}~\ref{fig:bat-age} para naman sa taga-Batangas. Ayon sa {\em Figure}~\ref{fig:lag-age}a, malaking bahagdan ng populasyon ang nakakabata (18--40 taon gulang), sa parehong kasarian, kesa sa mga matatanda ($>$ 40 taon gulang). Malinaw na ipinapakita ng pigura na habang tumataas ang gulang, mas kumakaunti ang sumasapi sa OSN, subalit isang kamangha-manghang katutuhanan ang ipinakita ng datos na mayroong mga kasapi ang $>60$ taong gulang na. Ipinakita naman ng mga {\em Figure}~\ref{fig:lag-age}b at~c na mas higit na nakararami ang mga walang asawa, sa parehong kasarian, na may gulang na $<30$ taon gulang. Dahil dito, masasabi natin na ang anyo ng OSN ng mga taga-Laguna ay mga kabataang ($<30$ taon gulang) walang asawa.

Ang {\em Figure}~\ref{fig:bat-age}a ay nagpapakita ng malaking bilang ng mga kasapi sa Facebook na taga-Batangas ang mga kabataan (18-40 taon gulang), sa parehong kasarian, kesa sa mga may gulang na $>40$ taon. Ang obserbasyong ito ay kapareho ng sa taga-Laguna: Mas kakaunti ang mga kasapi ng OSN ang nakatatanda, subalit mayroong mga kasaping $>60$ taong gulang. Ipinakita naman ng {\em Figure}~\ref{fig:bat-age}b at~c na maraming bilang ng OSN ang mga kabataang walang asawa ($<30$ taong gulang). Mula sa ating diskurso ng mga datos mula sa taga-Laguna at taga-Batangas, masasabi natin na ang mga populasyon ng mga eto sa Facebook ay magkatulad at ang mga populasyong ito ay binubuo ng mga walang asawang kabataan.

\subsection{Pagpili ng ka-{\em Friend}}
\begin{figure}[hbt]
\centering\epsfig{file=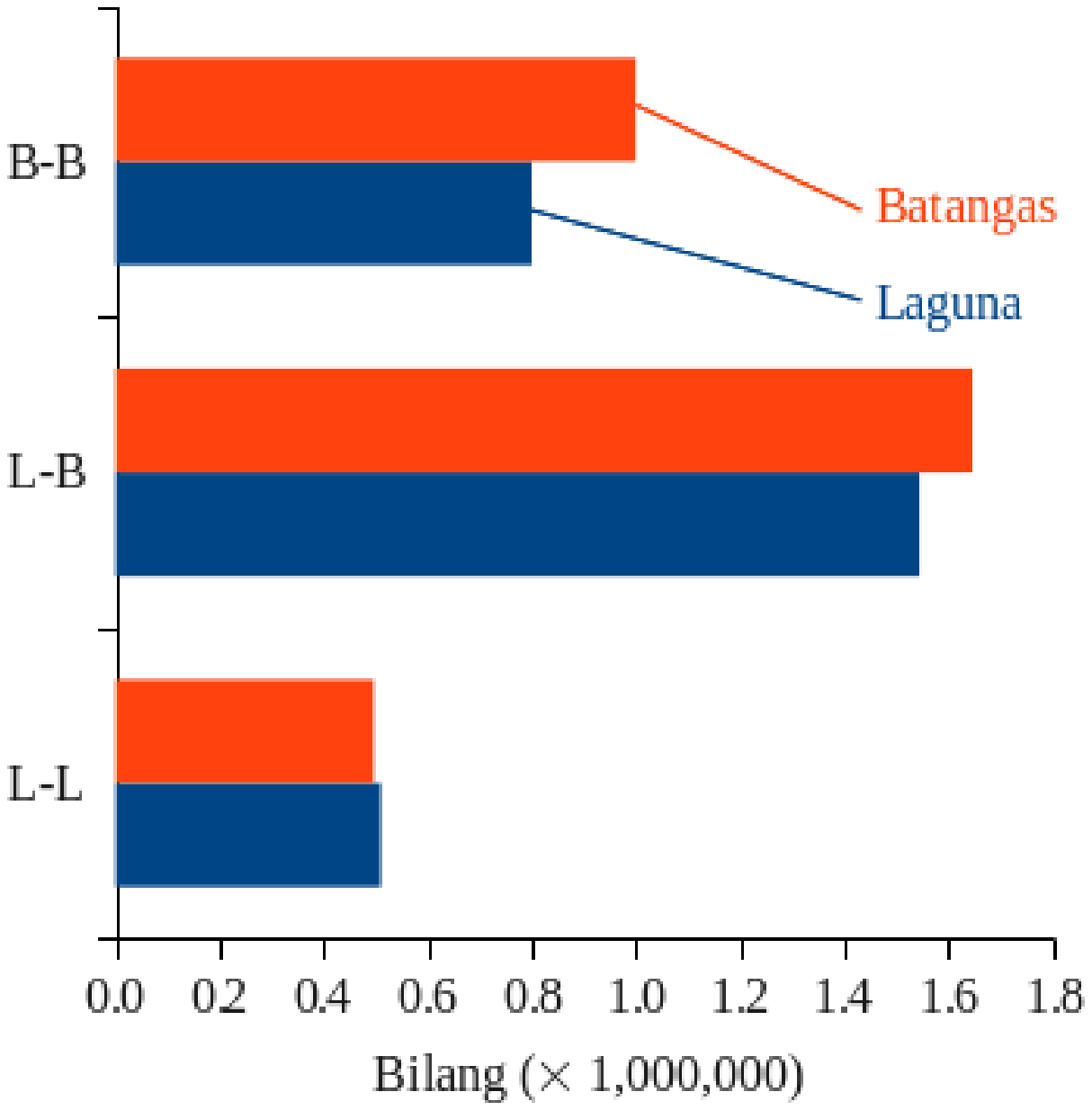, width=3in}
\caption{Bilang ng pagkakaibigan sa pamamagitan ng kasarian sa parehong lalawigan.}\label{fig:friend-gender}
\end{figure}

Ang anyo ng pagpili ng ka-{\em friend}, base sa kasarian, ay aming nakalap galing sa {\em database}. Ang mga bilang ng pagkakaibigan sa pagitan ng dalawang lalaki, dalawang babae, at isang babae at isang lalaki ay awtomatikong nakalap ng {\em web robot}. Ang mga bilang na ito ay ipinapakita sa {\em Figure}~\ref{fig:friend-gender} para sa mga taga-Laguna at mga taga-Batangas. Ang pigura ay malinaw na nagpapakita na ang mga pagkakaibigan ng magkakaibang kasarian ay mas madalas na nangyayari kesa doon sa mga magkakapareho. Kaya, maari nating sabihin na ang mga kasapi ng OSN sa Facebook ay mas pinipiling maging ka-{\em friend} ang mga kasaping iba ang kasarian kesa sa kanila. Ang anyong ito ay nag papahiwatig ng {\em heterophilly} sa pagtangi ng kaibigan base sa kasarian.

\begin{figure}[hbt]
\centering\epsfig{file=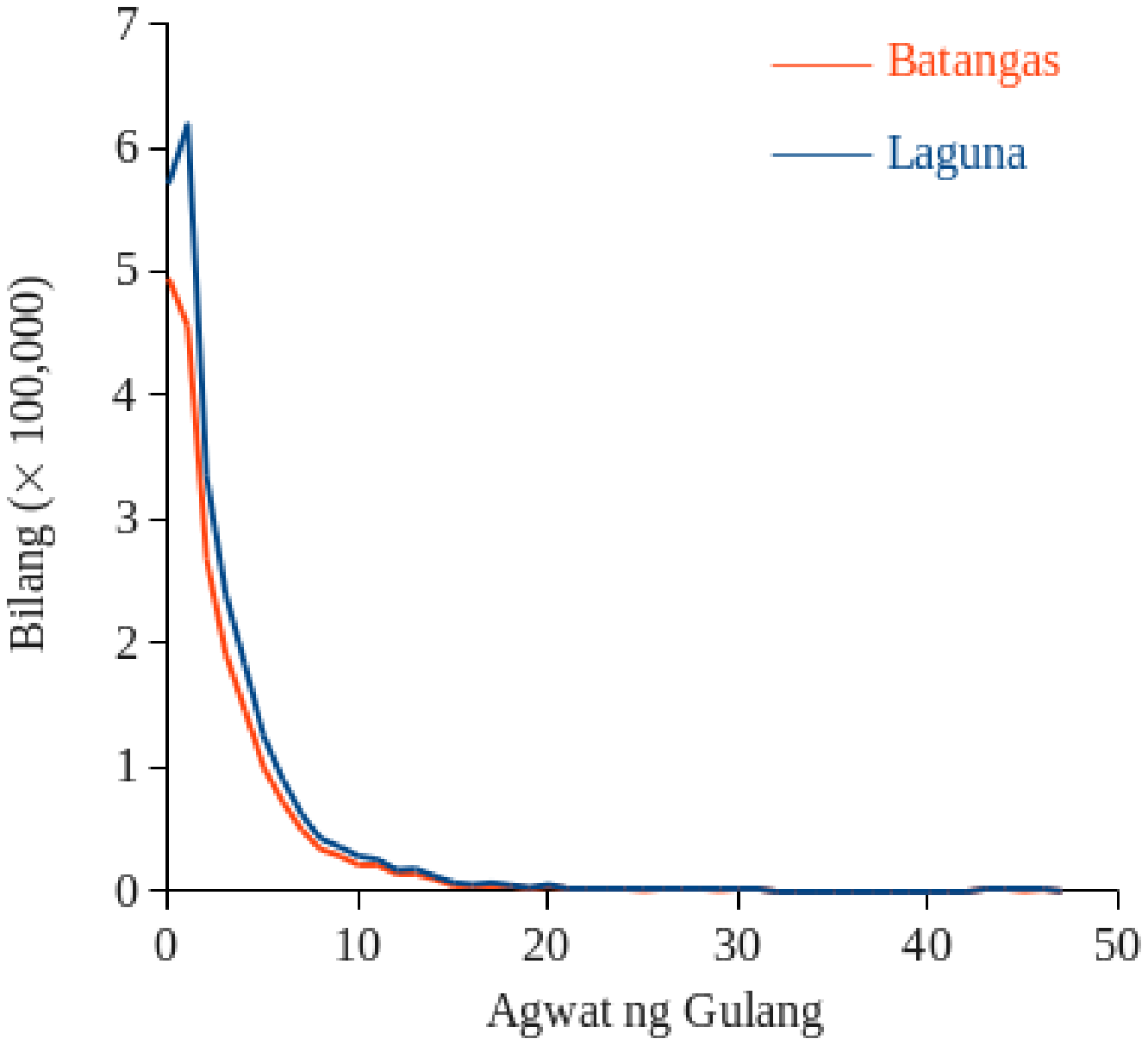, width=3in}
\caption{Bilang ng pagkakaibigan sa pamamagitan ng agwat ng gulang sa parehong lalawigan.}\label{fig:friend-age}
\end{figure}

Ang {\em Figure}~\ref{fig:friend-age} ay nagpapakita ng bilang ng mga ka-{\em friend} sa pagitan ng alin man sa dalawang kasapi na may agwat ng gulang mula~0 hanggang~47 na taon. Ang pigura ay nagpapakita na ang mga pagkakaibigan ng mga kasapi, mula sa dalawang lalawigan, na agwat ng gulang na hindi tataas sa 10 taon ay mas nangyayaring madalas kesa doon sa ang agwat ay mas mataas sa 10 taon. Ang anyong ito ay nagpapakita ng {\em homophilly} sa pagpili ng kaibigan sa pamamagitan ng agwat ng taon.

\begin{figure}[hbt]
\centering\epsfig{file=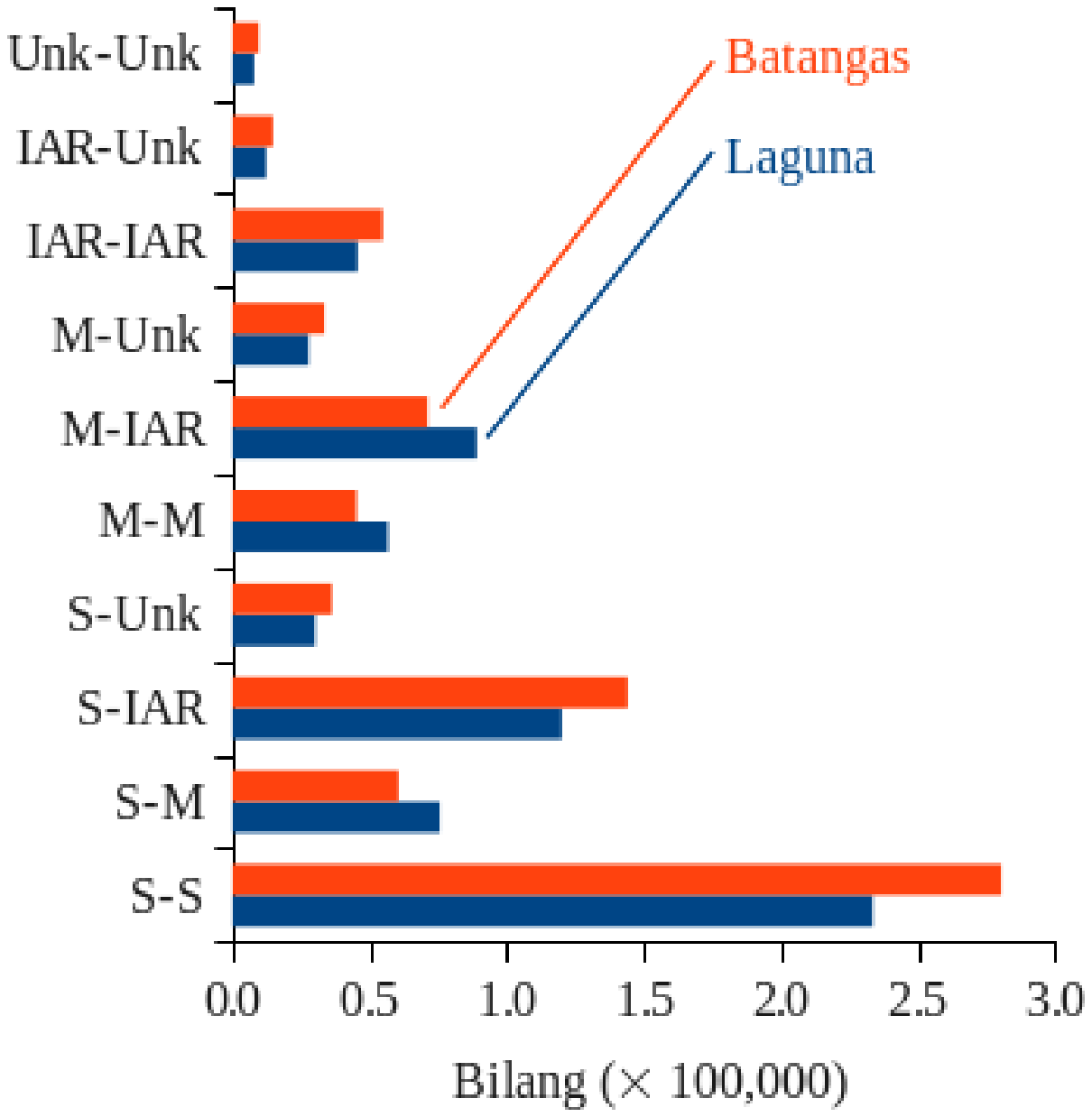, width=3in}
\caption{Bilang ng pagkakaibigan sa pamamagitan ng estadong pang-sibil sa parehong lalawigan.}\label{fig:friend-status}
\end{figure}

Sa {\em Figure}~\ref{fig:friend-status}, ipinapakita ang bilang ng mga pagkakaibigan mula sa Laguna at Batangas sa pamamagitan ng estadong pang-sibil. Ang pigura ay nagpapakita na mas maraming nabuong ka-{\em friend} sa pagitan ng dalawang walang asawa (isang anyo ng {\em homophilly}), kasunod ang bilang ng mga naging mag-kaibigan sa pagitan ng walang asawa at IAR, at sa pagitan ng may asawa at IAR (parehong anyo ng {\em heterophilly}). Ang mga kasaping kasalukuyang may ka-relasyon ay nakikipag-kaibigan pa rin sa mga kasaping kasalukuyang may ka-relasyon na ({\em homophilly}). Gayon din, ang mga kasaping may asawa na ay nakiki-{\em friend} din sa mga may asawa na ({\em homophilly}). kailangan naming banggitin dito na ang unang dalawang pangngungusap ay hindi dapat bigyan ng negatibong kahulugan dahil sa ang mga ito ay nagsasalamin lamang ng mga buod ng datos. Hindi alam ng datos kung ang pakikipag-kaibigan ng isang may asawa sa ibang kasapi na may asawa na, halimbawa, ay kung ang dalawang ito ang talagang mag-asawa.

\subsection{Istraktura ng OSN}
Ang resulta ng aming {\em path analysis} sa mga OSN ng mga taga-Laguna at ng mga taga-Batangas ay ang mga sumusunod: (1)~Ang pinakamahaba at pamantayang {\em path length} ng mga taga-Laguna ay 11 at 4.6, ayun sa pagkakasunud-sunod; at (2)~ang sa mga taga-Batangas naman ay 12 at 4.4. Ito ay nagpapahiwatig na ang isang kasapi~$v_i$ na taga-Laguna ay maaabot ang isa pang kasapi~$v_k$, sa pamamagitan ng pamantayang 4.6 na {\em padrino}, at si~$v_i$ ay mayroong garantiyang maabot si~$v_k$ sa loob lamang ng 11~na mga {\em padrino}. Kung sina $v_i$ at $v_k$ ay mga taga-Batangas, si~$v_i$ ay maabot si~$v_k$ sa pagitan ng pamantayang 4.4 na {\em padrino}, at siya ay may garantiyang maabot si~$v_k$ sa pagitan ng hindi na hihigit pa sa 12~na {\em padrino}. Ang mga datos na ito ay nagpapahiwatig na ang dalawang OSN ay parehong {\em small-world}. Sila ay mas maliit pa sa pamayanang inilahad nina Travers at Milgram~\citep{travers69} na may pamantayang 6~na {\em padrino}. Sa ibang teksto, ang batayang ito ay tinawag na ``{\em six degrees of separation}.''

\begin{figure}[hbt]
\centering\epsfig{file=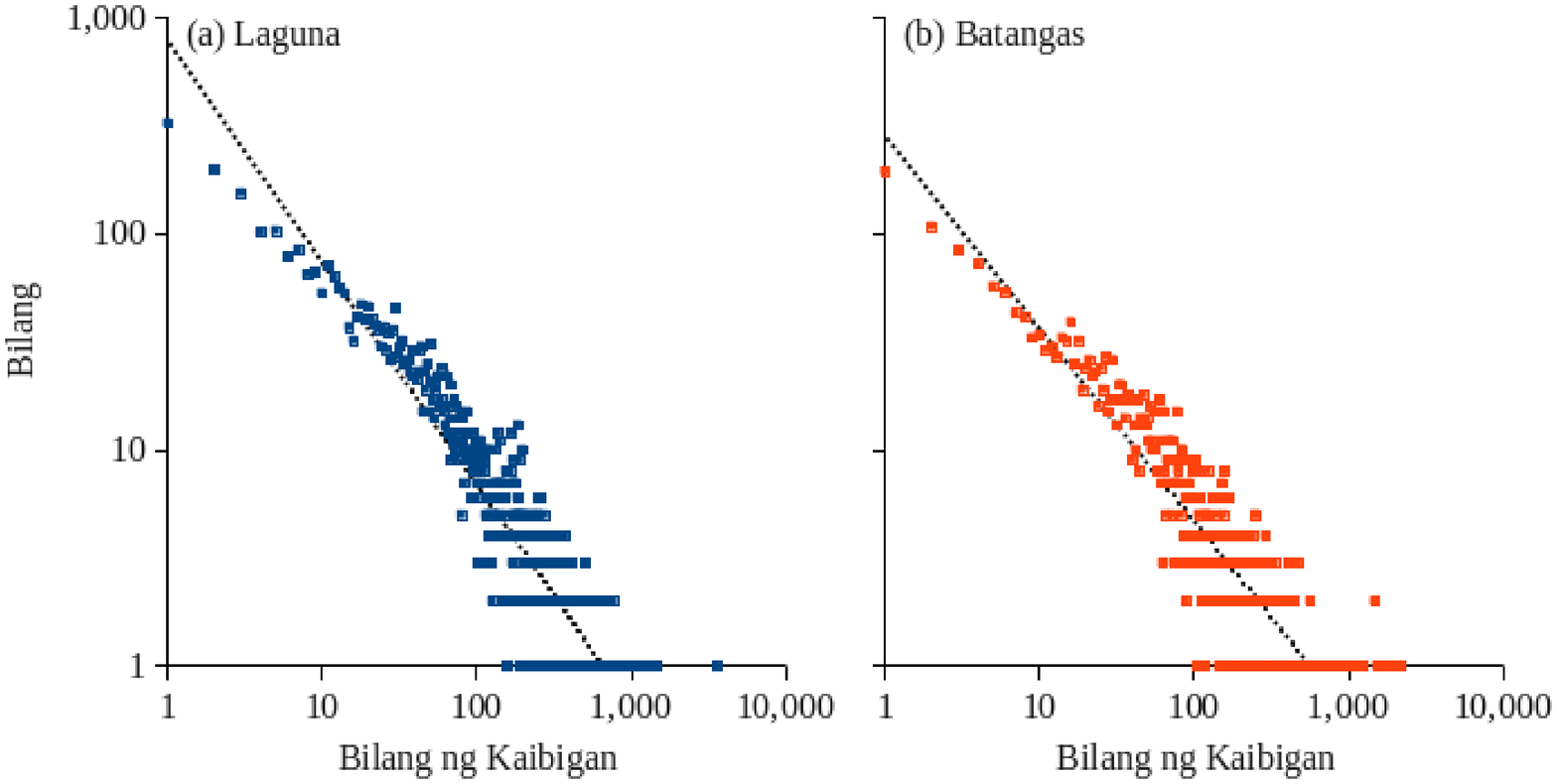, width=5in}
\caption{Ang log-log na pigura sa pagitan ng bilang at bilang ng kaibigan sa (a) Laguna at sa (b) Batangas. Ang mga bughaw na mga parisukat ay mga datos at ang linya ay ang {\em power law regression line}, isang pagpapatunay na ang mga datos ay sumusunod sa {\em power law distribution} .}\label{fig:scale-free}
\end{figure}

Ang {\em Figure}~\ref{fig:scale-free} ay nagpapakita ng kung gaano kalimit ang mga bilang ng kaibigan para sa (a)~mga taga-Laguna at (b)~mga taga-Batangas. Ang mga pigura ay parehong nagsasaad ng {\em power law distribution}, kaya ang mga OSN ng mga taga-Laguna at mga taga-Batangas ay parehong masasabing {\em scale-free}. Ang pagkakaroon ng {\em mabigat na buntot} ng dalawang pigura ay nagsasaad na maraming mga kasapi ang pumapapel bilang mga pusod o sentro ng kani-kanilang mga OSN. Ang mga sentrong ito ay mga kasaping may malaki at malawak na empluwensiya sa ibang kasapi ng OSN, isang impormasyon na maaring magamit sa mga kampanyang pang-politikal at pang-komersyo, gayon din sa mga pagpapakalat ng mga negatibong datos. Halimbawa, ang mga tsismis ay madaling kakalat sa OSN kung ang mga ito ay magsisimula sa mga sentrong ito.

\section{Buod at Pagpapatibay}\label{summary}
Ang diskursong ito ay nagpakita ng paggamit ng isang {\em web robot} para awtomatikong makalap ang mga datos ng mga kasapi ng Facebook na mula sa Laguna at Batangas. Ang mga datos na ito ay sinuri para maihambing ang mga OSN na galing sa dalawang lalawigan. Mula sa mga pagsusuri, napag-alaman na ang mga lalawigang ito ay magkatulad sa mga sumusunod:
\begin{enumerate}
	\item Ang populasyon ay dominado ng mga kabataang babae na hindi hihigit sa 30 taong gulang at walang asawa (o ka-relasyon);
	\item Kapansin-pansin ang homophily sa pagpili ng ka-{\em friend} sa pamamagitan ng gulang (mas madali silang maging magkaibigan kung hindi mas mataas mula sa lima hanggang sampung taon ang agwat ng mga gulang nila); 
	\item Heterophily naman ang mapapansin sa pagpili ng ka-{\em friend} kung ang pagbabasehan ay ang kasarian (mas maraming relasyon ang nabubuo sa pagitan ng isang babae at isang lalaki kesa sa pagitan ng dalawang babae o dalawang lalaki);
	\item Mas maraming mga pagkakaibigan ang nabuo sa pagitan ng dalawang walang asawa, ng dalawang may asawa, o ng dalawang IAR ({\em homophilly}), subalit ang iba ay nagpapakita na ng {\em heterophilly} katulad ng sa pagitan ng may asawa at IAR, at walang asawa at IAR;
	\item Ang istraktura ng OSN ng parehong pamayanan ay masasabing {\em small-world} at {\em scale-free}; at
	\item Ang mga OSN ay mayroong mga kasaping malaki at malawak ang empluwensiya kumpara sa ibang kasapi.
\end{enumerate}

\section*{Pasasalamat}

Ang pagsasaliksik na ito ay may tulong na pananalapi mula sa Surian ng Agham Pang-kompyuter ng \UPLB. Ito ay isinagawa sa pagitan ng ika-11 ng Hulyo hanggang ika-6 ng Agosto 2013 gamit ang 12--{\em node computer cluster} sa {\em Research Collaboratory for Advanced Intelligent System} ng Surian ng Agham Pang-kompyuter. Ang pagsasaliksik na ito ay halaw sa mga programang {\em Structural Characterization and Temporal Dynamics of Various Natural, Social and Artificial Networks in the Philippines} at {\em Evaluating Approaches for Modeling and Simulating the Structure and Dynamics of Artificial Societies} na pinamumunuan ni Prof. Pabico.


\bibliography{facebook}
\bibliographystyle{unsrtnat}

\end{document}